\documentclass[twocolumn,prd,preprintnumbers,superscriptaddress,nofootinbib,showpacs]{revtex4}

\usepackage{amsmath}
\usepackage{amsfonts}
\usepackage{amssymb}
\usepackage{graphicx}
\usepackage{color}
\usepackage{hyperref}
\usepackage{cleveref}
\usepackage{booktabs}
\usepackage{color}
\usepackage{braket}
\usepackage{mathtools}
\usepackage[rightcaption]{sidecap}
\usepackage{subfigure}
\usepackage{tensor}

\definecolor{vividviolet}{rgb}{0.62, 0.0, 1.0}
\definecolor{amaranth}{rgb}{0.9, 0.17, 0.31}
\definecolor{palatinateblue}{rgb}{0.15, 0.23, 0.89}
\definecolor{brightpink}{rgb}{1.0, 0.0, 0.5}
\definecolor{cornflowerblue}{rgb}{0.39, 0.58, 0.93}
\definecolor{deepcarminepink}{rgb}{0.94, 0.19, 0.22}
\definecolor{radicalred}{rgb}{1.0, 0.21, 0.37}

\hypersetup{ linktoc=all,
    colorlinks, linkcolor={palatinateblue},
    citecolor={brightpink}, urlcolor={amaranth}
}

\newcommand{\be}{\begin{equation}}
\newcommand{\ee}{\end{equation}}
\newcommand{\bs}{\begin{split}}
\newcommand{\bea}{\begin{eqnarray}}
\newcommand{\eea}{\end{eqnarray}}

\newcommand{\bes}{\begin{subequations}}
\newcommand{\ees}{\end{subequations}}

\begin{document}

\title{Geometric corrections to cosmological entanglement}

\author{Alessio Belfiglio}
\email{alessio.belfiglio@unicam.it}
\affiliation{Scuola di Scienze e Tecnologie, Universit\`a di Camerino, Via Madonna delle Carceri 9, 62032 Camerino, Italy.}
\affiliation{Istituto Nazionale di Fisica Nucleare, Sezione di Perugia, Via Alessandro Pascoli 23c, 06123 Perugia, Italy.}

\author{Orlando Luongo}
\email{orlando.luongo@unicam.it}
\affiliation{Scuola di Scienze e Tecnologie, Universit\`a di Camerino, Via Madonna delle Carceri 9, 62032 Camerino, Italy.}
\affiliation{Dipartimento di Matematica, Universit\`a di Pisa, Largo B. Pontecorvo 5, Pisa, 56127, Italy.}
\affiliation{NNLOT, Al-Farabi Kazakh National University, Al-Farabi av. 71, 050040 Almaty, Kazakhstan.}

\author{Stefano Mancini}
\email{stefano.mancini@unicam.it}
\affiliation{Scuola di Scienze e Tecnologie, Universit\`a di Camerino, Via Madonna delle Carceri 9, 62032 Camerino, Italy.}
\affiliation{Istituto Nazionale di Fisica Nucleare, Sezione di Perugia, Via Alessandro Pascoli 23c, 06123 Perugia, Italy.}

\begin{abstract}
We investigate entanglement production by inhomogeneous perturbations over a homogeneous and isotropic cosmic background, demonstrating that the interplay between quantum and geometric effects can have relevant consequences on entanglement entropy, with respect to homogeneous scenarios. To do so, we focus on a conformally coupled scalar field and discuss how geometric production of scalar particles leads to entanglement. Perturbatively, at first order we find oscillations in entropy correction, whereas at second order the underlying geometry induces mode-mixing on entanglement production. We thus quantify entanglement solely due to geometrical contribution and compare our outcomes with previous findings. We characterize the geometric contribution through geometric (quasi)-particles, interpreted as dark matter candidates.
\end{abstract}


\maketitle
\tableofcontents


\section{Overview} \label{sez1}

Ascertaining entanglement in curved spacetime remains an outstanding issue of theoretical physics \cite{1,2,3,4,5,6,7}. In this respect, the Universe expansion, predicted by Einstein's field equations, can plausibly generate particles \cite{8,9,10}, giving rise to entanglement as due to quantum field evolution in curved spacetime \cite{11,12,13,14,15,16,17}. These models mostly assume a homogeneous cosmological Friedmann-Robertson-Walker (FRW) background, so the corresponding entanglement singles out only particle pairs with opposite momenta.  Thus, a natural step to generalize this framework is to include inhomogeneities, that play a significant role in the early Universe. Inhomogeneous perturbations departing from a genuine FRW leads to  pair-creation probability depending on local geometric quantities, related to gravitational overdensities, that also modify entanglement production, as gravity becomes locally ‘lumpy’.
We here unveil that inhomogeneities lead to non-negligible particle creation with mode-mixing, providing relevant consequences on mode dependence over entanglement measures. We demonstrate that spacetime geometry can modify entanglement production once  ``geometric particles" are naturally produced as consequence of a given (perturbed) cosmological background. We thus critically revise previous non-conclusive efforts, showing how entanglement might significantly differ. To do so, we select an asymptotically flat scale factor \cite{9,18}, and  compute the entanglement entropy up to second order in perturbations. Our goals are: $a)$ we investigate  perturbatively the quantum and geometric effects on entanglement generation due to inhomogeneities, $b)$ we underline oscillations in particle production at first order, due to the concurrent effects of expansion and geometry, $c)$ we argue how a purely geometric contribution arises at second order and characterize how entanglement changes under the effects of \emph{geometric particles}, $d)$ we propose a potentially exhaustive interpretation to explain dark matter's nature from geometric particles, $e)$ we  come up with geometric particles as \emph{geometric quasi-particles}, partially fueling the dark matter energy momentum budget of the Universe.

The manuscript is outlined as follows. In Sec. \ref{sez2},  we present the model and describe our perturbative approach to inhomogeneities. In Sec. \ref{sez4}  we discuss particle production up to second order. In Sec. \ref{sez5} we study entanglement generation both for massive and massless particles. In Sec. \ref{sez6} we draw our conclusions. Natural units,  $G=\hbar=c=1$, are here used.


\section{Scalar particles in perturbed spacetimes}\label{sez2}

We require scalar particles with given mass $m$ non-minimally coupled to spacetime scalar curvature, namely $R$, perturbing with small inhomogeneities the spatially-flat FRW metric, i.e.,  $g_{\mu \nu}\simeq a^2(\tau) \left( \eta_{\mu \nu}+h_{\mu \nu} \right)$. Here $a(\tau)$ is the scale factor, $\tau$ the conformal time, $\eta_{\mu \nu}$ the Minkowski metric  and $h_{\mu \nu}$ the small inhomogeneities,  $\lvert h_{\mu \nu} \rvert \ll 1$. The scalar field Lagrangian,
\be \label{1}
\mathcal{L}= \frac{1}{2} \sqrt{-g}\left[ g^{\mu \nu}\phi_{, \mu} \phi_{,\nu}-(m^2+\xi R)\phi^2\right],
\ee
can be then perturbed up to the first order \cite{19,20} as $\mathcal{L}\simeq \mathcal{L}^{(0)}+\mathcal{L}^{(1)}+\mathcal O(h^2)$. Assuming conformal coupling, namely $\xi=1/6$, the free modes are analytical, enabling the expansion field in terms of \textit{in}, and/or \textit{out}, positive frequency modes by
\be
\begin{aligned} \label{2}
\hat{\phi}({\bf x}, \tau)&= \frac{1}{a} \int \frac{d^3k}{(2 \pi)^{3/2}} \frac{1}{\sqrt{2 E_{\text{in}}(k)}} \\
&\ \ \ \ \ \ \ \ \ \times \left[ f_k^{\text{in}}({\bf x}, \tau) \hat{a}_{\text{in}}({\bf k}) + f_k^{\text{in}*}({\bf x}, \tau) \hat{a}^\dagger_{\text{in}}({\bf k})  \right],
\end{aligned}
\ee
where  $E_{\text{in}}(k)=\sqrt{\lvert {\bf k} \rvert^2+m^2 a^2(-\infty)}$. The modes $f_k$ can be expressed as $f_k({\bf x}, \tau)= f_k(\tau) e^{i {\bf k} \cdot {\bf x}}$, with $f_k(\tau)$ satisfying
\be \label{4}
\ddot{f}_k(\tau)+ \left[ \lvert {\bf k} \rvert^2+ m^2 a^2(\tau) \right] f_k(\tau)=0.
\ee
Since we now perturb the system, we can work in the interaction picture and express the system final state as
\be \label{5}
\lim_{\tau \rightarrow + \infty} \lvert \Psi \rangle= \mathcal{N} \left[ \lvert 0, \text{in} \rangle + \sum_n \frac{1}{n!} \lvert n, \text{in} \rangle \langle n, \text{in} \rvert S^{(1)} \lvert 0, \text{in} \rangle \right],
\ee
where $S^{(1)}=i \hat{T} \int \mathcal{L}^{(1)}\  d^4x$  is the first order S-matrix. Here the vector $\lvert n \rangle$ symbolizes any state containing $n $ particles, while $\mathcal{N}$ is a normalizing factor.  The interaction Lagrangian is quadratic in the field and its derivatives, thus particles are produced in pairs up to first order in $h_{\mu \nu}$. The probability amplitude for pair creation is given by the S-matrix element $
S^{(1)}_{kp}= \langle {\bf k} {\bf p} \lvert S^{(1)} \rvert 0 \rangle \notag = i \int \langle {\bf k} {\bf p} \lvert \hat{T} \mathcal{L}^{(1)} \rvert 0 \rangle \ d^4x$,
where all the states are intended as \textit{in} states. This element gives the transition from the vacuum state $\lvert 0 \rangle$ to a two-particle state $\lvert {\bf k} {\bf p} \rangle$, with momenta ${\bf k}$ and ${\bf p}$. An asymptotically flat spacetime is needful to guarantee vacuum uniqueness, so we single out the widely-adopted scale factor  \cite{18} $a^2(\tau)=A+B \tanh \rho\tau$, with $A$ and $B$ parameters controlling  Universe's volume, while $\rho$ is related to Universe's expansion rapidity. We suppose the perturbation is not negligible on a finite time interval, namely $\tau \in [\tau_i,\tau_f]$, with $\tau_i$ and $\tau_f$ negative and $ \lvert \tau_i \rvert, \lvert \tau_f \rvert \gg 1$. Accordingly,
\be \label{7}
h_{\mu \nu}=\begin{cases}
h_{\mu \nu},\ \ \text{if}\   \tau_i < \tau < \tau_f\\
0,\ \ \ \ \ \ \text{otherwise}.
\end{cases}
\ee
Within this interval, we can safely assume $f_k^{\text{in}}(\tau) \simeq e^{-i E_{\text{in}} \tau}$, provided the expansion is sufficiently fast. Accordingly, in synchronous gauge, $h_{0 \nu}=0$, we can derive the pair creation amplitude \cite{20}
\begin{widetext}
\be \label{8}
S_{kp}^{(1)}= \frac{i\tilde{h}_{\mu \nu}(k+p)}{2(2 \pi)^3 \sqrt{E_{\text{in}}(k)E_{\text{in}}(p)}} \left[ k^{\mu} p^{\nu} -\frac{1}{6} (k+p)^{\mu} (k+p)^{\nu} -\frac{1}{12} \eta^{\mu \nu} (k+p)^\sigma (k+p)_\sigma- \frac{1}{2} \eta^{\mu \nu} m^2 a^2(\tau \rightarrow -\infty) \right].
\ee
\end{widetext}


From the amplitude \eqref{8}, we can numerically get the amount of created particles. So, it behooves us to thoroughly specify  $h_{\mu \nu}$ components. In the synchronous gauge, the metric components $g_{00}$ and $g_{0i}$ are unperturbed and so
\be \label{9}
ds^2= a^2(\tau) \left[ d\tau^2-(\delta_{ij}+h_{ij})dx^i dx^j \right],
\ee
where $i,j=1,2,3$.
We focus here on scalar modes, portrayed by $h$ and by the traceless part $h_{ij}^{\parallel}$.
Scalar perturbations are easier to compute in \textit{conformal Newtonian gauge}, as the metric is diagonal and one can recognize the gravitational potential, $\psi$, in Newtonian limit  \cite{21}. So, we have
\be \label{10}
ds^2= a^2(\tau) \left[ (1+2\psi)d\tau^2-(1-2\phi) dx^id x_i \right],
\ee
where the second scalar, $\phi$, is required only if the energy-momentum tensor contains a non-vanishing traceless and longitudinal component. At a first glace, we may set $\psi=\phi$  and consider nearly Newtonian perturbation source, i.e., $\psi=-M/r$, where $M$ is the mass which generates the perturbation and $r$ the radial coordinate. The corresponding scalar perturbation in synchronous gauge, $h_{ij}=h/3+ h_{ij}^\parallel$, can be derived easily following Ref. \cite{21}, with the prescription $\dot{a}/a \simeq 0$ in  $[\tau_i,\tau_f]$. The details are reported in Appendix \ref{app0}, giving
\begin{widetext}
\be \label{11}
h_{\mu \nu} (x)= -M \begin{pmatrix} 0 & 0 & 0 & 0 \\
0 & \left[ \frac{2}{r} + \left(  \frac{3x^2-r^2}{r^5}-\frac{4 \pi}{3} \delta({\bf r}) \right) \tau^2\right] & 3 \tau^2 \left(\frac{xy}{r^5} \right)  & 3 \tau^2 \left(\frac{xz}{r^5}\right) \\ 0 & 3 \tau^2 \left(\frac{xy}{r^5}\right) & \left[ \frac{2}{r} + \left(  \frac{3y^2-r^2}{r^5}- \frac{4 \pi}{3} \delta({\bf r}) \right) \tau^2 \right] & 3 \tau^2 \left(\frac{yz}{r^5}\right) \\ 0 & 3 \tau^2 \left(\frac{xz}{r^5}\right) & 3 \tau^2 \left(\frac{yz}{r^5}\right) & \left[ \frac{2}{r} + \left( \frac{3z^2-r^2}{r^5}- \frac{4 \pi}{3} \delta({\bf r}) \right) \tau^2 \right]
\end{pmatrix}.
\ee
\end{widetext}
We remark that our assumption of vanishing perturbation outside $[\tau_i,\tau_f]$ can be interpreted in terms of particle backreaction on the spacetime structure \cite{22,23,24,25,26,27}. In fact, it has been pointed out that the reaction of particle creation back on the gravitational field is able to reduce the creation rate, damping out initial perturbations on timescales of the order of Planck's time. Even if our work is based on the usual external field approximation, backreaction may in principle justify the transient nature of the perturbation.
Bearing $h_{\mu\nu}$ in mind, we can now derive particle production, as we report below.


\section{Particle production}\label{sez4}

At first  perturbation order, pair production is due to the combined effect of the expansion and inhomogeneities, whereas at second order the two contributions give instead distinct effects. Hence, \emph{the production rate is nonzero even if the homogeneous background does not produce particles}, i.e., as the \textit{in} and \text{out} vacua are identical.

At first order, the asymptotic \textit{out} state  \eqref{5} takes the form
\be \label{12}
\lvert \Psi \rangle_{out}\equiv \lim_{\tau \rightarrow + \infty} \lvert \Psi \rangle = \mathcal{N} \left( \lvert 0, \text{in} \rangle + \frac{1}{2} S_{kp}^{(1)}\  \lvert {\bf k} {\bf p}, \text{in} \rangle  \right),
\ee
with $\mathcal{N}=1+\mathcal O(h^2)$  a normalization factor. Introducing   Bogoliubov transformations, that relate \textit{in} and \textit{out} ladder operators between them \cite{13}, we get the first order number density \cite{19,20}
\be \label{13}
n^{(1)}(k,p)= (2 \pi a_f)^{-3} \delta^3({\bf k}+{\bf p}) \text{Re}\bigg[ S^{(1)}_{kp} \big(\alpha_k^* \beta_k+\alpha_p^*\beta_p\big) \bigg],
\ee
where  $a_f \equiv a(\tau \rightarrow + \infty)$ and $\alpha_k,\beta_k$ are the Bogoliubov coefficients. It is clear from Eq. \eqref{13} that particles are created in pairs with opposite momenta. Introducing a generic lower bound cut-off scale $r_{\text{min}}$, regularizing the integral with $\propto e^{-r}$, to avoid divergences at infinity, we can compute the Fourier transform of \eqref{11} and obtain
\be \label{14}
\tilde{h}_{ij}(2E_{\text{in}},0)= -8\pi M e^{-r_{\text{min}}} (1+r_{\text{min}})  \int_{\tau_i}^{\tau_f} e^{2iE_{\text{in}}\tau}\ d\tau\  \delta_{ij},
\ee
Eq. \eqref{14} can be used to derive the probability amplitude \eqref{8} and then the number density \eqref{13}. We notice that \emph{massless particles are not produced at first order, since $\lvert \beta_k \rvert=0$}. At second order the number density reads instead
\be \label{15}
n^{(2)}(k,p)= \mathcal{N}^2\  (2\pi a_f)^{-3}\   \lvert S^{(1)}_{kp}\rvert^2\ \left(\lvert \beta_k \rvert^2+ \lvert \beta_p \rvert^2 +1  \right),
\ee
where we  exploited the normalization condition $\lvert \alpha_q \rvert^2-\lvert \beta_q \rvert^2=1$ ($q=k,p$) and $\mathcal{N}$ is  straightforwardly computed from $\langle \Psi \rvert \Psi \rangle=1$.
We notice that \emph{at second order there is a purely geometric contribution}, i.e., particles are produced even if $\beta_q=0$. As perturbations live where Universe's expansion is almost negligible, the  probability $\lvert S^{(1)}_{kp} \rvert^2$ can be computed explicitly. We separately discuss the massive and massless cases.

\paragraph{Massive particles.}

For slow expansion rate, or Minkowskian background,  pair production probability for massive conformally coupled particles is given by 
\begin{align} \label{16}
W^{(1)}&= \int d^3k\ d^3p\  \lvert S^{(1)}_{kp} \rvert^2 \notag \\
&=\int \frac{d^4 q}{(2 \pi)^4} \frac{\theta(q^0) \theta(q^2-4m^2)}{960 \pi} \left( 1-\frac{4m^2}{q^2}\right)^{1/2} \notag \\
&\ \ \ \ \times \bigg[ \tilde{C}_{\mu \nu \rho \sigma}(q) \tilde{C}^{\mu \nu \rho \sigma}(-q) \left( 1-\frac{4m^2}{q^2}\right)^2  \notag \\
&\ \ \ \ \ \ \ \ \ \ \ +\frac{20}{3} \frac{m^4}{q^4} \tilde{R}(q) \tilde{R}(-q)\bigg],
\end{align}
where in the last equality we  introduced the four momentum $q$, i.e.,
$q=(q^0, {\bf q} )=(k^0+p^0, {\bf k}+{\bf p})$. In Eq. \eqref{16}, $\tilde{C}_{\mu \nu \rho \sigma}(q)$ and $\tilde{R}(q)$ are the Fourier transforms of the Weyl tensor and the scalar curvature, respectively.  These quantities are derived in Appendix \ref{appA}, as functions of the perturbation tensor $h_{\mu \nu}$. Accordingly, pair production probability can be written in terms of local geometric quantities. In order to compute the  $h_{\mu \nu}(x)$ Fourier transform, we assume that  the particle momenta are along the $z$ direction, without losing generality, having
\be \label{17}
\tilde{h}_{\mu \nu} (q)= \int d\tau\  e^{iq^0\tau} \int d^3r\  e^{i q_z  r \cos \theta} e^{-r}\  h_{\mu \nu}(x),
\ee
where $q_z$ is the total momentum.

Within the framework of massive particles, we can speculate on the fact that the interaction between curvature and scalar field implies a promising scenario toward the existence of geometric quasi-particles of dark matter. Gravitational dark matter production has been recently explored, focusing both on scalar and vector candidates, (see e.g. \cite{10} for a recent review). We here conjecture the geometric production enables quasi-particle candidates for dark matter. The scalar field $\phi$ can be treated as a suitable dark matter contributor \cite{28,29} and, as no interactions are provided with external fields, a benchmark mass $m=1$ MeV \cite{28,30,31} can be assumed for numerical studies.

\paragraph{Massless particles.} Massless particles can be produced at second order in perturbation, since inhomogeneities break the conformal symmetry of the theory. If $m=0$, the probability \eqref{16} only depends on deviations from conformal flatness, quantified by the Weyl tensor. This result is valid in any FRW spacetime, i.e., it is not strictly request a slowly expanding background\footnote{This is due to the fact that the Weyl tensor $C^{\mu}_{\hphantom{\mu} \nu \rho \sigma}(x)$ is invariant under conformal transformations.}.


\section{Geometric entanglement} \label{sez5}

Up to first order particles are produced in pairs with opposite momenta, i.e., there is no mode-mixing. We write the \textit{in} vacuum in a Schmidt decomposition of \textit{out} states \cite{11}
\be \label{18}
\lvert 0_k; 0_{-k} \rangle_{\text{in}}=  \sum_{n=0}^{\infty} c_n \lvert n_k; n_{-k} \rangle_{\text{out}},
\ee
where $c_n$ is a normalization constant and $n_k$ labels the number of excitations in the field mode $k$ as seen by an inertial observer in the \textit{out} region. The \textit{in} state containing one pair,
\be \label{19}
\lvert 1_k;1_{-k} \rangle_{\text{in}} = \hat{a}_{\text{in}}^\dagger({\bf k}) \hat{a}_{\text{in}}^\dagger(-{\bf k}) \lvert 0_k; 0_{-k}\rangle_{\text{in}},
\ee
can be written in the \textit{out} region exploiting the inverse Bogoliubov transformations, as described in Appendix \ref{appB}. To evaluate the entanglement amount, we consider the first order density operator $\rho^{\text{out}}_{k,-k}= \lvert \Psi \rangle_{\text{out}} \langle \Psi \rvert$ and Eq. \eqref{12} with ${\bf p}=-{\bf k}$. Performing a partial trace over one subsystem, we obtain the reduced density operator
\be \label{20}
\rho^{\text{out}}_k=\text{Tr}_{-k} \bigg(\rho^{\text{out}}_{k,-k}\bigg).
\ee
The coefficients $c_n$ are derived from the normalization $\langle \Psi \rvert \Psi \rangle_{\text{out}}=1$. Introducing the quantity $\gamma= \lvert \beta_k^*/\alpha_k \rvert^2$, after some algebra we arrive to the final expression for the density operator
\begin{align} \label{21}
\rho^{\text{out}}_k&=\frac{(1-\gamma)^2}{1-\gamma+(1+\gamma)\text{Re}\big( S_{k,-k}^{(1)} \alpha_k^* \beta_k \big)}\notag \\
&\ \ \times \sum_{n=0}^\infty \gamma^n \bigg( 1+ \text{Re}\big( S_{k,-k}^{(1)} \alpha_k^* \beta_k \big) (2n+1) \bigg) \lvert n \rangle_k \langle n \rvert.
\end{align}

\begin{figure}
    \centering
    \includegraphics[scale=0.63]{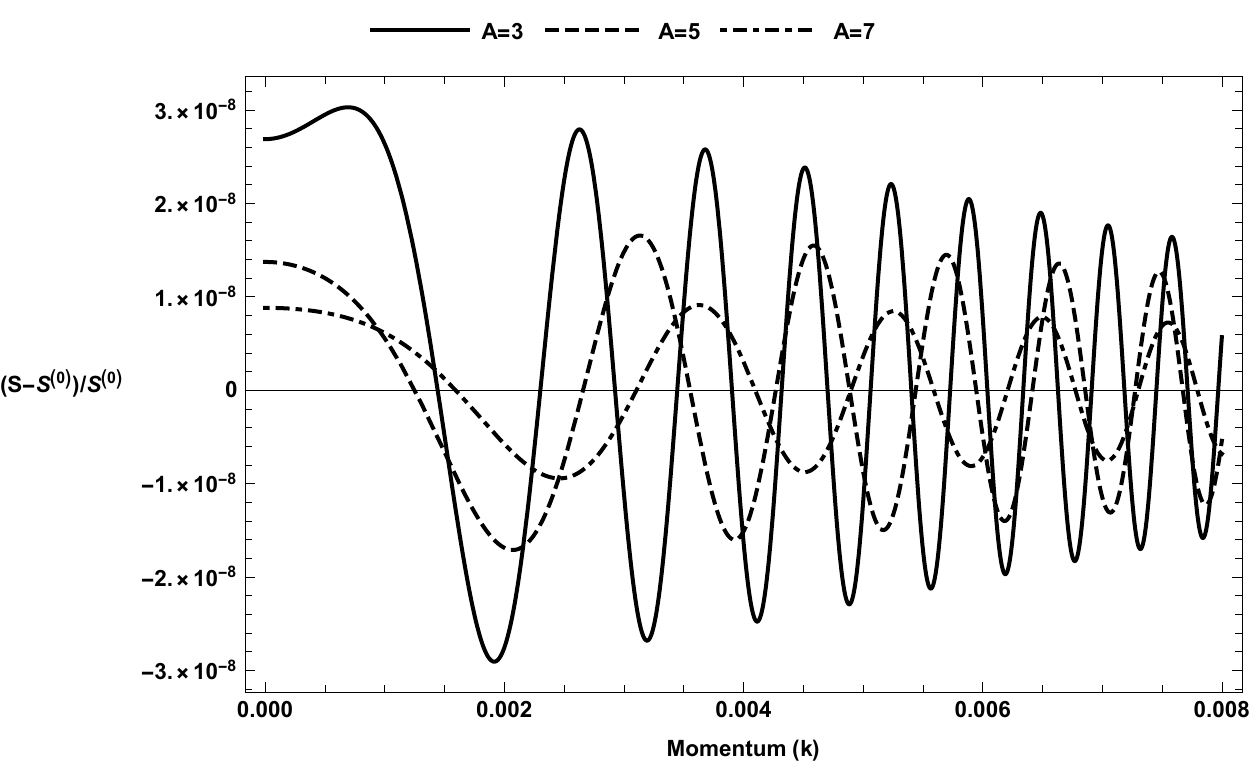}
    \caption{Entropy shift  as function of the momentum $k$, for different values of the parameter $A$. The other parameters are $B=2$, $\rho=1$, $m=0.01$, $n=40$, $M=10^{-5}$, $r_{\text{min}}=5$, $\tau_i=-10^4$ and $\Delta \tau \equiv \tau_f- \tau_i=100$.}
    \label{fig1}
\end{figure}

The corresponding von Neumann entropy is then
\begin{align} \label{22}
S(\rho_k^{\text{out}})&= S^{(0)}(k)+S^{(1)}(k) =-\text{Tr}(\rho_k^{\text{out}} \log_2 \rho_k^{\text{out}})\notag \\
&= - \sum_{n=0}^\infty \lambda_n \log_2 \lambda_n,
\end{align}
where  $\lambda_n$ are the $\rho_k^{\text{out}}$ eigenvalues, say
\begin{align} \label{23}
\lambda_n&= \frac{(1-\gamma)^2}{1-\gamma+(1+\gamma)\text{Re}\big( S_{k,-k}^{(1)} \alpha_k^* \beta_k \big)} \notag \\[6pt]
&\ \ \ \ \times \gamma^n\left( 1+ \text{Re}\big( S_{k,-k}^{(1)} \alpha_k^* \beta_k \big) (2n+1) \right).
\end{align}

In Fig. \ref{fig1} we show the \emph{entropy shift}, $(S-S^{(0)})/S^{(0)}$, for given values of $A$. The perturbation parameters are chosen so that $\lvert h_{\mu \nu} (x) \rvert \ll 1$, with the time parameters sufficiently large to fulfill the requirements of Sec. \ref{sez2}. Remarkably,  the first order contribution turns out to be quite small, whereas entanglement entropy oscillates. This means that \emph{kinds of  expansion and spacetime geometry can, in principle, decrease the amount of entanglement}, turning to be more relevant for small Universe volumes, i.e., small $A$. On the other hand,  the correction is larger at small momenta, due to the fact that at first order particles are mainly produced as $k$ is small. We expect first order corrections to be more relevant if the hypothesis of spherically symmetric perturbation is released.

We now turn to second order, where particles are produced in pairs with generic momenta $k$ and $p$, with a robust purely-geometric contribution. The corresponding entanglement generation is summarized below.

\begin{figure}[ht!]
    \centering
    \includegraphics[scale=0.63]{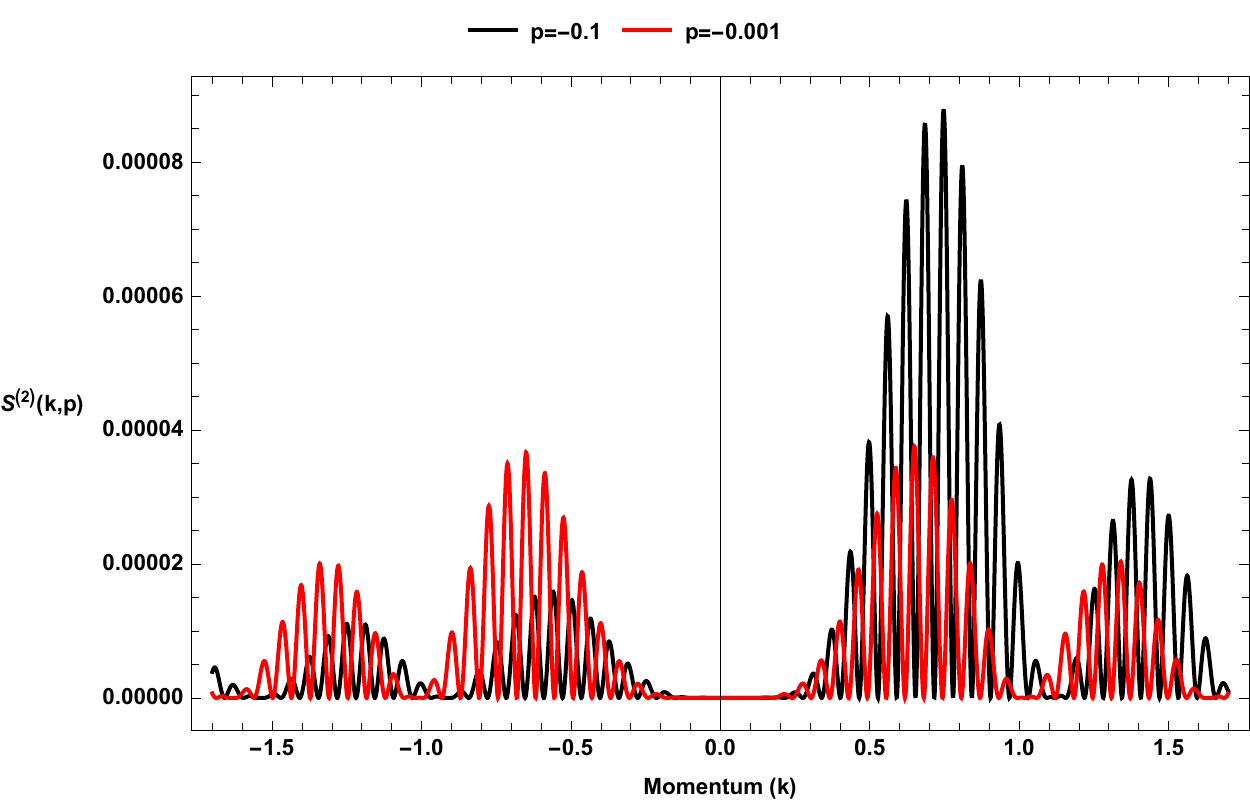}
    \caption{Entanglement entropy $S^{(2)}(k,p)$ as function of the particle momentum $k$, with $p=-0.1$ and $p=-0.001$. The other parameters are the same of Fig. \ref{fig1}, with $A=3$.}
    \label{fig2}
\end{figure}

\paragraph{Massive particles.} Here,  $n^{(2)}(k,p)$ is mainly settled by $\propto \lvert C_{\mu \nu \rho \sigma}(q) \rvert^2$ and $\propto \lvert R(q) \rvert^2$, as evident from Eq. \eqref{16}. It is then possible to set $\beta_k=\beta_p=0$, neglecting expansion effects, getting the state \eqref{12}  in terms of the \textit{out} basis as
\be \label{24}
\lvert \Psi \rangle= \mathcal{N} \left( \lvert 0_k ; 0_p \rangle + \frac{1}{2} S^{(1)}_{kp} \lvert 1_k; 1_p \rangle \right),
\ee
where we  remove the \textit{out} subscript since the \textit{in} and \textit{out} vacua coincide if the $\beta$ coefficients vanish. Eq. \eqref{24} shows up a bipartite pure state. In order to quantify the corresponding entanglement entropy, we trace out the "$p$" or "$k$" contribution. Accordingly, we are left with the following reduced density operator
\begin{align} \label{25}
\rho_k
= \mathcal{N}^2 \left( \lvert 0 \rangle_k \langle 0 \rvert+ \frac{1}{4} \lvert S_{kp}^{(1)} \rvert^2 \lvert 1 \rangle_k \langle 1 \rvert  \right),
\end{align}
where the probability of pair production $\lvert S^{(1)}_{kp} \rvert^2$ is given by Eq. \eqref{16}.
The subsystem entropy, following from Eq. \eqref{25}, is plotted in Fig. \ref{fig2} as function of  $k$.
Inhomogeneous perturbations break space translation symmetry, thus linear momentum is no longer conserved in particle creation processes. Accordingly, \emph{second order entropy is characterized by notable mode-mixing}.  This is an impressive property of geometric cosmological entanglement, never put forward.  We also notice that second order corrections to entanglement are typically larger than first order ones. As anticipated, this effect is related to our choice of a spherically symmetric perturbation, which makes the first order contribution negligible.

\begin{figure}
    \centering
    \includegraphics[scale=0.63]{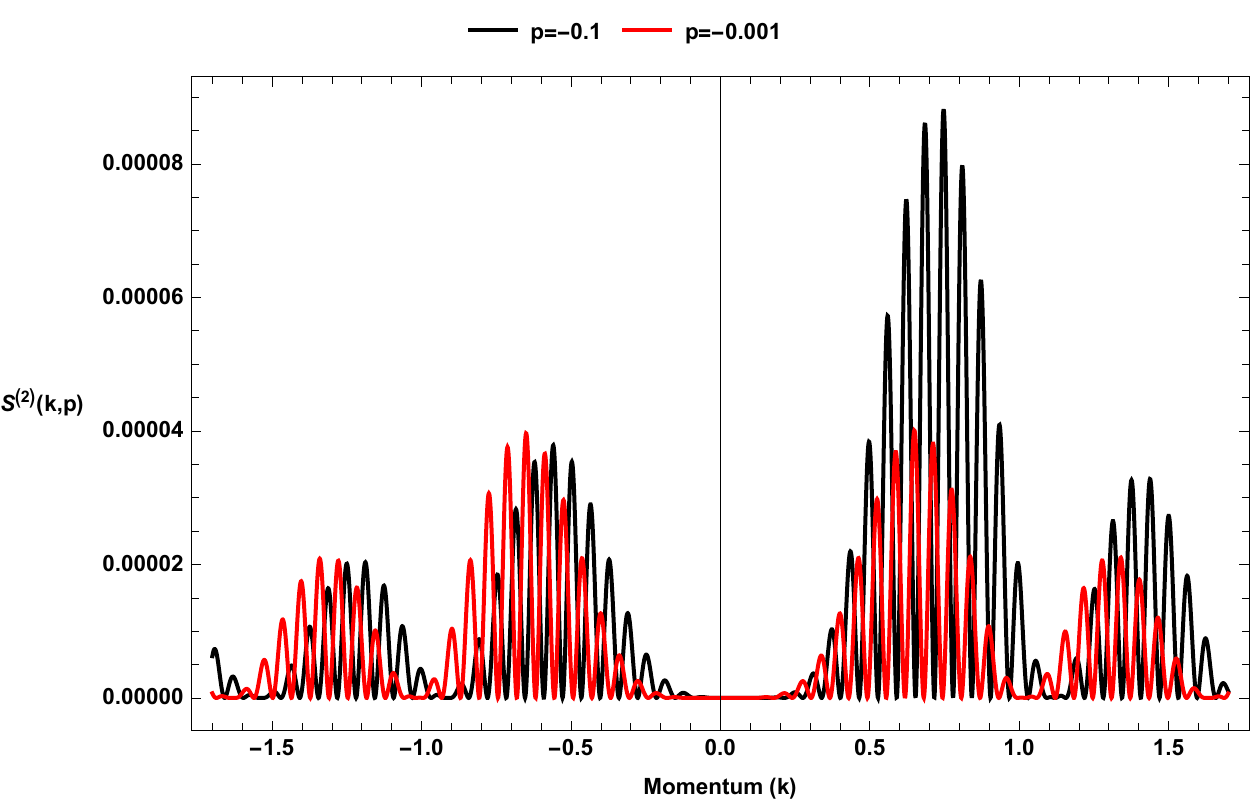}
    \caption{Entanglement entropy $S^{(2)}(k,p)$ as function of the momentum $k$ with $p=-0.1$ and $p=-0.001$. The other parameters are the same as in Fig. \ref{fig1}, with $A=3$.}
    \label{fig3}
\end{figure}

\paragraph{Massless particles.}
Here, $\beta_k=\beta_p=0$ is naturally fulfilled, due to the conformal symmetry. Starting from the state in Eq. \eqref{24}, the reduced density operator takes again the form \eqref{25} and  pair creation probability is determined by the Weyl tensor only. In Fig. \ref{fig3}, we display the entanglement entropy as function of the momentum $k$, assuming the same parameters as in the massive case.
The amount and mode dependence of entanglement closely resemble the massive case, thus showing that geometric effects on entanglement production turn out to be similar in both the aforementioned cases.


\section{Outlooks} \label{sez6}

We computed the entanglement amount in curved spacetime as due to inhomogeneous perturbations over a homogeneous FRW and confronted our findings with previous results developed in the literature. In particular, we demonstrated cosmological entanglement can be due to geometric contributions only. We then expect the matching between quantum and geometric effects drastically alter the entanglement measures. From the one hand we showed at first perturbative order oscillations in entropy corrections occurred, while at second order a non-negligible entropy, featured by non-vanishing mode-mixing, arises in entanglement generation.  To interpret this scheme, we noticed geometrical dark matter can be reviewed as a suitable benchmark scenario where \emph{dark matter emerges under the form of quasi-particles}. We proposed $\phi$ as a suitable dark matter contributor since no interactions are provided with external fields, suggesting a possible mass value around $m=1$ MeV, in agreement with recent observations. This paves the way for further investigations deepening the interconnections between the notions of entanglement and dark matter. According to our findings, we also expect geometry to exhibit relevant consequences on entanglement extraction protocols, e.g. entanglement harvesting. Consequently, this would open novel strategies to deduce cosmological parameters, to interpret dark matter under a new promising way, or to depict primordial quantum gravity stages.

\acknowledgments
The authors express their gratitude to Roberto Giambò and Peter K. S. Dunsby for useful discussions. OL acknowledges the support of the Ministry of Education and Science of the Republic of Kazakhstan, Grant IRN AP08052311.

\appendix

\section{Scalar perturbations in the synchronous and conformal Newtonian gauges} \label{app0}

We here briefly recall the gauge transformation between the synchronous and conformal Newtonian gauges  \cite{21}. We focus on scalar perturbations only. Let us consider a general coordinate transformation from a system $x^\mu$ to another $\hat{x}^\mu$
\be \label{S1}
x^\mu \rightarrow \hat{x}^\mu=x^\mu+d^\mu (x^\nu).
\ee
We write the time and the spatial parts separately as
\begin{align}
&\hat{x}^0= x^0 + \alpha({\bf x},\tau) \label{S2} \\
&\hat{{\bf x}}= {\bf x}+ \vec{\nabla} \beta({\bf x}, \tau)+ {\boldsymbol \epsilon}({\bf x}, \tau),\ \ \ \ \ \ \ \vec{\nabla} \cdot {\boldsymbol \epsilon}=0, \label{S3}
\end{align}
where the vector $d$ has been divided into a longitudinal component $\vec{\nabla} \beta$ and a transverse component $\vec{\epsilon}$. \\
Let $\hat{x}^\mu$ denote the synchronous coordinates and $x^\mu$ the conformal Newtonian coordinates, with $\hat{x}^\mu=x^\mu+d^\mu$. It is easy to see that
\begin{align}
& \alpha({\bf x}, \tau)=\dot{\beta}({\bf x}, \tau), \label{S4} \\
& \epsilon_i({\bf x}, \tau)=\epsilon_i({\bf x}), \label{S5} \\
& h_{ij}^\parallel({\bf x}, \tau)=-2 \bigg( \partial_i \partial_j-\frac{1}{3} \delta_{ij} \nabla^2 \bigg) \beta({\bf x}, \tau), \label{S6} \\
& \partial_i \epsilon_j+ \partial_j \epsilon_i=0. \label{S7}
\end{align}
and
\begin{align}
    & \psi({\bf x}, \tau)= -\ddot{\beta}({\bf x}, \tau)- \frac{\dot{a}}{a} \dot{\beta}({\bf x}, \tau), \label{S8}\\
    & \phi({\bf x}, \tau)= +\frac{1}{6} h({\bf x}, \tau)+ \frac{1}{3} \nabla^2 \beta({\bf x}, \tau)+ \frac{\dot{a}}{a} \dot{\beta}({\bf x}, \tau) \label{S9}.
\end{align}
Assuming an asymptotically flat spacetime as described in Sec. \ref{sez2}, we can safely conclude that $\dot{a}/a \simeq 0$ in the region $[t_i,t_f]$. Accordingly, \eqref{S8} would give
\be \label{S10}
\beta(r,\tau) \simeq +\frac{M}{2r} \tau^2.
\ee
Subtracting now \eqref{S9} from \eqref{S8}, with the assumption  $\psi=\phi$, we obtain
\be \label{S11}
h(r, \tau)=-6 \ddot{\beta}- 2 \nabla^2 \beta= -2M \left[ \frac{3}{r}-2\pi\tau^2 \delta({\bf r}) \right].
\ee
Finally, from \eqref{S6} we get
\be \label{S12}
h_{ij}^\parallel (r, \tau)= -M\tau^2 \left[\partial_i \partial_j \left( \frac{1}{r} \right) - \frac{1}{3} \delta_{ij} \left(-4\pi \delta({\bf r}) \right)  \right].
\ee
Recalling the well known result
\be \label{S13}
\partial_i \partial_j \left( \frac{1}{r} \right)= \frac{3x_i x_j-r^2 \delta_{ij}}{r^5}- \frac{4 \pi}{3} \delta({\bf r}) \delta_{ij}
\ee
and the expression for scalar perturbations in synchronous gauge, we finally arrive at the tensor $h_{\mu \nu}$, namely Eq. \eqref{11}.


\section{Curvatures and Weyl tensor in linearized gravity} \label{appA}

As discussed in Sec. \ref{sez4}, pair production probability at second order depends on local geometric quantities. Here, we recall the main results from linearized gravity, which are useful in order to derive the probability of pair creation \eqref{16}. Starting from the perturbed Minkowski metric
\be \label{A1}
g_{\mu \nu}=\eta_{\mu \nu}+h_{\mu \nu}(x),
\ee
the connection coefficients are
\be \label{A2}
\Gamma^{\rho}_{\hphantom{\rho} \mu \nu}= \frac{1}{2} \eta^{\rho \sigma} \left( \partial_{\mu} h_{\nu \sigma}+\partial_{\nu} h_{\mu \sigma}-\partial_{\sigma} h_{\mu \nu} \right).
\ee
Accordingly, the first order Riemann curvature is
\begin{align} \label{A3}
R_{\mu \nu \rho \sigma}&= \eta_{\mu \lambda} \left( \partial_\sigma \Gamma^{\lambda}_{\hphantom{\lambda} \nu \rho}- \partial_\rho \Gamma^{\lambda}_{\hphantom{\lambda} \nu \sigma} \right) \notag \\
&=  \partial_{\rho} \partial_{[\nu} h_{\mu] \sigma}+ \partial_{\sigma} \partial_{[\mu} h_{\nu] \rho},
\end{align}
where square brackets denote antisymmetrization, as usual. The Ricci curvature follows as
\be \label{A4}
R_{\mu \nu}= R^\rho_{\mu \rho \nu}= \frac{1}{2}\partial^\rho \partial_{\rho} h_{\mu \nu} -\partial^{\rho} \partial_{(\mu} h_{\nu) \rho} + \frac{1}{2}\partial_\mu \partial_\nu h,
\ee
and the Ricci scalar as
\be \label{A5}
R= \eta^{\mu \nu} R_{\mu \nu}= \partial^\mu \partial_\mu h- \partial^\mu \partial^\nu h_{\mu \nu}.
\ee
Introducing now the Fourier transform of the perturbation
\be \label{A6}
\tilde{h}_{\mu \nu}= \int d^4q\ e^{i q x} h_{\mu \nu}(x),
\ee
it is straightforward to obtain
\be \label{A7}
\tilde{R}(q)= q^{\mu} q^{\nu} \tilde{h}_{\mu \nu}(q)-q^2 \tilde{h}(q).
\ee
Assuming a real perturbation, one finds
\begin{widetext}
\begin{align}
    &\lvert R(q) \rvert^2= R(q) R(-q)= q_{\mu} q_{\nu} q_{\rho} q_{\sigma} h^{\mu \nu}(q) h^{\rho \sigma}(-q)+q^4 h(q) h(-q)-q^2 q_{\mu} q_{\nu} \left[ h^{\mu \nu}(q) h(-q)+ h^{\mu \nu}(q) h(-q) \right], \label{A8} \\
    & \lvert R_{\mu \nu}(q) \rvert^2= \frac{1}{2} q_{\mu} q_{\nu} q_{\rho} q_{\sigma} h^{\mu \rho}(q) h^{\nu \sigma}(-q)-\frac{q^2}{2} q_{\rho} q^\mu h_{\mu \nu} (q) h^{\nu \rho}(-q)+\frac{1}{4} q^4 h^{\mu \nu}(q) h_{\mu \nu}(-q) \notag \\
    &\ \ \ \ \ \ \ \ \ \ \ \ \ \  -\frac{1}{4} q^2 q_{\mu} q_{\rho} \left[ h^{\mu \rho}(q)h(-q)+h^{\mu \rho}(-q) h(q) \right], \label{A9} \\
    &\lvert R_{\mu \nu \rho \sigma} \rvert^2= \frac{1}{4} q^4 h^{\mu \nu}(q) h_{\mu \nu}(-q)-2 q^2 q_{\mu} q^{\nu} h_{\nu \sigma}(q) h^{\mu \sigma}(-q)+ q_{\mu} q_{\nu} q_{\rho} q_{\sigma} h^{\mu \rho}(q) h^{\nu \sigma}(-q). \label{A10}
\end{align}
\end{widetext}
From Eqs. \eqref{A8}-\eqref{A10} it can be shown that
\be \label{A11}
\lvert R_{\mu \nu \rho \sigma} \rvert^2-4\lvert R_{\mu \nu} \rvert^2+ \lvert R \rvert^2=0.
\ee
In four dimensions, the Weyl conformal tensor takes the form
\be \label{A12}
C_{\mu \nu \rho \sigma}= R_{\mu \nu \rho \sigma}+ \frac{1}{3}R g_{\mu [ \sigma}g_{\rho] \nu}+\left( g_{\nu [ \rho} R_{\sigma ] \mu}-  g_{\mu [ \rho} R_{\sigma ] \nu}\right),
\ee
that gives
\begin{align} \label{A13}
\lvert C_{\mu \nu \rho \sigma} (q) \rvert^2&= C_{\mu \nu \rho \sigma}(q) C^{\mu \nu \rho \sigma}(-q) \notag \\
&= \lvert R_{\mu \nu \rho \sigma} \rvert^2-2 \lvert R_{\mu \nu} \rvert^2+ \frac{1}{3} \lvert R \rvert^2.
\end{align}
Now, exploiting  Eq. \eqref{A11} we can rewrite Eq. \eqref{A12} as
\be \label{A14}
\lvert C_{\mu \nu \rho \sigma} (q) \rvert^2 = 2 \lvert R_{\mu \nu} \rvert^2- \frac{2}{3} \lvert R \rvert^2.
\ee
The probability of pair production at second order in the perturbation, both for massive and massless particles, can be computed starting from the result \eqref{A14}, as discussed in \cite{18,19}.


\section{First order density operator} \label{appB}

Here we derive the explicit form of the reduced density operator \eqref{20}, which is required in order to quantify first order corrections to entanglement entropy. In the \textit{out} region, the two-particle state \eqref{19} reads
\begin{widetext}
\begin{align} \label{B1}
\lvert 1_k;1_{-k} \rangle_{\text{in}} &= \big(\alpha_k^*\hat{a}_\text{out}^\dagger({\bf k}) + \beta_k \hat{a}_{\text{out}}(-{\bf k}) \big) \big(\alpha_k^*\hat{a}_\text{out}^\dagger(-{\bf k}) + \beta_k \hat{a}_{\text{out}}({\bf k}) \big)  \sum_{n=0}^{\infty} c_n \lvert n_k; n_{-k} \rangle_{\text{out}} \notag \\
&= (\alpha_k^*)^2 \sum_{n=0}^{\infty} (n+1) c_n\  \lvert n+1; n+1 \rangle_{\text{out}} + \alpha_k^*\beta_k \sum_{n=0}^{\infty} n c_n\  \lvert n_k; n_{_k} \rangle_{\text{out}} \notag \\
& + \alpha_k^*\beta_k \sum_{n=0}^{\infty} (n+1) c_n\  \lvert n_k; n_{-k} \rangle_{\text{out}} + \beta_k^2 \sum_{n=0}^{\infty}n c_n\ \lvert n-1;n-1 \rangle_{\text{out}},
\end{align}
where we have exploited the Bogoliubov transformations relating asymptotic ladder operators, which has the general form \cite{13}
\begin{align}
    & \hat{a}_{\text{out}}({\bf k})= \alpha_k^* \hat{a}_{\text{in}}({\bf k})- \beta^*_k \hat{a}^\dagger_{\text{in}}(-{\bf k}), \label{B2}  \\
    & \hat{a}_{\text{in}}({\bf k})= \alpha_k \hat{a}_{\text{out}}({\bf k})+ \beta^*_k \hat{a}^\dagger_{\text{out}}(-{\bf k}). \label{B3}
\end{align}
From the state \eqref{12} and the expansion \eqref{18} we can then write the density operator $\rho^{\text{(out)}}_{k,-k}= \lvert \Psi \rangle_{\text{out}} \langle \Psi \rvert$ up to first order, and performing a partial trace over antiparticles, we are left with
\begin{align} \label{B4}
\rho^{\text{out}}_k&=\text{Tr}_{-k} \bigg(\rho^{\text{out}}_{k,-k}\bigg) \notag \\
&=\sum_{n=0}^{\infty} \lvert c_n \rvert^2 \lvert n \rangle_k \langle n \rvert \notag \\
&\ \ \ +\frac{1}{2}S_{k,-k}^{(1)} \alpha_k^* \beta_k \bigg[ \sum_{n=0}^{\infty} (n+1) \lvert c_n \rvert^2\ \lvert n \rangle_k \langle n \rvert + \sum_{n=0}^{\infty} n \lvert c_n \rvert^2 \ \lvert n \rangle_k \langle n \rvert \bigg] \notag \\
&\ \ \ +\frac{1}{2} S_{k,-k}^{(1)*}  \alpha_k \beta_k^* \bigg[  \sum_{n=0}^{\infty} (n+1) \lvert c_n \rvert^2\ \lvert n \rangle_k \langle n \rvert +  \sum_{n=0}^{\infty} n \lvert c_n \rvert^2 \ \lvert n \rangle_k \langle n \rvert \bigg].
\end{align}

\end{widetext}
\end{document}